\documentclass[aps,prl,onecolumn,groupedaddress]{revtex4-1}
\usepackage{natbib}

\usepackage{mhchem}
\usepackage{times,mathptmx}
\usepackage{amsmath} 
\usepackage{amssymb} 
\usepackage{graphicx} 
\usepackage{epstopdf}
\usepackage{xcolor}
\usepackage{subcaption}

\begin{document}

\newcommand{\tcol}[1]{\textcolor{red}{#1}}
\newcommand{\cl}{C_1(h,t)}
\newcommand{\cd}{C_2(h,t)}
\newcommand{\cp}{\zeta(h,t)}

\title{Sedimentation and diffusion of passive particles in suspensions of swimming \emph{Escherichia coli}}


\author{Jaspreet Singh$^{1,*}$, Alison E. Patteson$^{1,*}$, Prashant K. Purohit$^{1}$, Paulo E. Arratia$^{1,**}$}


\address{$^{1}$Department of Mechanical Engineering \& Applied Mechanics,
	University of Pennsylvania, Philadelphia, PA 19104.\\
$^{*}$Equal contribution \\
$^{**}$Corresponding Author \\ Email address: parratia@seas.upenn.edu}


\date{\today}

\begin{abstract}
Sedimentation in active fluids has come into focus due to the ubiquity of swimming micro-organisms in natural and artificial environments. Here, we experimentally investigate sedimentation of passive particles in water containing various concentrations of the bacterium {\it E. coli}. Results show that the presence of living bacteria reduces the velocity of the sedimentation front even in the dilute regime, where the sedimentation velocity is expected to be independent of particle concentration. Bacteria increase the effective diffusion coefficient of the passive particles, which determines the width of the sedimentation front. For higher bacteria concentration, we find the development of two sedimentation fronts due to bacterial death. A model in which an advection-diffusion equation describing the settling of particles under gravity is coupled to the population dynamics of the bacteria seems to capture the experimental trends relatively well.
\end{abstract}


\maketitle

\section{Introduction}

The settling of organic and inorganic matter in fluids plays an important role in many technological and natural processes. Proper dispersion of particulates in liquids is essential to the production of foodstuff, paints, biofuels, and plastics, while sedimentation of biological matter regulates planktonic organisms' position relative to light and foraging strategies and is a key part of the ocean carbon cycle (i.e. ocean's biological pump) that transports carbon from the surface oceans to depth. Recently, there has been much interest in the sedimentation of active particles, which are usually defined as self-propelling particles (living or synthetic) that inject energy, generate mechanical stresses, and create flows within the fluid medium \cite{marchetti2013}. These particles can drive the fluid out of equilibrium (even in the absence of external forcing) and lead to many interesting phenomena such as collective behavior \cite{marchetti2013,sriram2017}, unusual viscosity \cite{lopez2015,lindner2013}, and an enhancement in particle diffusivity \cite{Wu2000,Chen2007,Leptos2009,Mino2011,Jepson2013} that depends anomalously on particle size \cite{Patteson2016}. Describing such active systems remains challenging, particularly under the effects of external forcing such as gravity \cite{Cates2008,Cates2009,Cates2010,PattesonOpinion}.

Recent studies have mainly focused on the \textit{steady-state} sedimentation of suspensions of active particles. Experiments with dilute active colloids such as phoretic particles found that density profiles at steady state decay exponentially with height yielding a sedimentation length that is larger than that expected for thermal equilibrium \cite{Palacci2010,Ginot2015}. This enhancement can be described by an effective activity-induced temperature that correlates with the particle's ability to self-propel and achieve larger diffusivities than from thermal fluctuations alone. These results agree relatively well with theory \cite{Cates2008,Cates2009} and simulations \cite{Cates2010, Tsao2014} for active particles that are either non-interacting \cite{Cates2008,Cates2009} or with limited hydrodynamic interactions \cite{Cates2010, Tsao2014}. 

Experiments with swimming micro-organisms paint a more detailed picture of sedimentation in active systems. For instance, under an external centrifugal field, \textit{Escherichia coli} (\textit{E}. \textit{coli}) fractionizes by motility so that fast-swimming bacteria swim throughout the sample and slow-swimming bacteria accumulate at the bottom; the resultant particle distribution matches a model of active colloids that possess a spectrum of effective temperatures \cite{Leonardo2013}. In the presence of extra-cellular polymers, bacteria can aggregate, enhancing sedimentation rates \cite{Poon2012}; however motile bacteria are more resistant to this aggregation than non-motile bacteria due to their enhanced diffusivity. In mixtures of swimming algae and passive particles, the steady-state sedimentation profile of passive particle is found to be described by an effective diffusivity that increases linearly with the concentration of swimming microbes \cite{Polin2016}. While the concept of effective temperatures and enhanced diffusivities have been useful in describing the steady-state sedimentation profiles of active systems, the transient unsteady evolution of the systems remains largely unknown. An important question is: How does a distribution of an initially homogenous mixture of active and passive particle suspension subject to gravity change over time?

In this manuscript, we investigate the sedimentation dynamics of active bacterial suspensions in experiments and using a simple model. The bacterium {\it E. coli}, a model organism widely used for motility and genetic studies \cite{berg2008}, are used to create active suspensions that are seeded with passive polystyrene (PS) colloidal particles. We observe these initially well-mixed suspensions as they settle over long periods of time (over 48 hours) and use image analysis techniques to track the evolving density profile and the spreading interface at the top of the sedimenting particle suspension (Fig. 1). We find that the presence of bacteria increases the macroscopic diffusivity of the passive particles (consistent with studies tracking the microscopic displacement of single particles \cite{Wu2000,Chen2007,Mino2011,Jepson2013,Patteson2016}), which in turn increases the width of the sedimentation front. Moreover, the presence of bacteria reduces the suspension sedimentation velocity, even in the dilute regime. This decrease in sedimentation velocity is significantly more than would be expected for increases in volume fraction due to the addition of bacteria, which swim force-free. At higher bacterial concentrations, the death of bacterial lead to the development of two sedimentation fronts, one due to passive particles and one due to inactive bacteria. We capture the main features of these fronts using an advection-diffusion equation coupled with population dynamics to account for bacterial death.

\section{Experimental Methods}

The experimental fluids are suspensions of swimming  {\it Escherichia coli} (wild-type K12 MG1655) and passive polystyrene particles in a buffer solution. The bacterium {\it E. coli} is a model organism for flagellated bacteria motility and achieves net propulsion by rotating their helical flagella at approximately 100 Hz, driving the cell body forward at speeds of $10$-$20$ $\mu$m/s \cite{berg2008}. The (time-averaged) flow generated by swimming \emph{E. coli} are well approximated by a force dipole that decays with the distance from cell body $r$ as 1/$r^2$\cite{Goldstein2011}. Here, bacteria are grown to saturation ($10^9$ cells/mL) in culture media (LB broth, Sigma-Aldrich). The saturated culture is gently cleaned by centrifugation and is suspended in buffer at concentration $c$ ranging from 0.75 to 7.5 $\times 10^9$ cells/mL. These concentrations are considered dilute, corresponding to volume fractions $\phi = c v_{\mathrm{b}} <$1\%, where $v_{\mathrm{b}}=$ 1.4 $\mu$m$^3$ is the  \emph{E. coli} body's volume~\cite{Jepson2013}. 

Polystyrene spheres (Sigma Aldrich) with a diameter $d$ of $2$ $\mu$m are used as passive particles. Polystyrene particles are cleaned by centrifugation and then resuspended in the buffer-bacterial suspension. Particle concentrations are dilute at $1.0 \times 10^8$ particles/mL, which corresponds to $0.14 \%$ volume fraction. Consistent with previous predictions and measurements on the concentration of bacteria ($\approx 10^{10}$ cells/mL) for the onset of collective motion \cite{Kasyap2014}, we do not observe any large scale collective behavior in these particle/bacteria suspensions.




\footnotetext{\textit{$^{a}$~Department of Mechanical Engineering \& Applied Mechanics,
		University of Pennsylvania, Philadelphia, PA 19104. E-mail: parratia@seas.upenn.edu}}


To image the sedimentation process, we introduce 1.5 mL of the experimental fluid into glass vials with a diameter of 8.3 mm and height of 16.6 mm, as shown schematically in Fig. \ref{Fig_1}a. The suspensions are gently mixed by hand with a pipette so that the particles are uniformly distributed at the start of the experiment ($t=0$ hr). The vials are capped and air volume (approximately {175} mm$^3$) remains inside them. In order to reduce the light diffraction from the round vials and to control temperature, the samples are placed in a cube-shaped water bath maintained at $T_{0}=22^{\circ}$C. Images are taken every 1 to 10 minutes for up to 7 days {with a Nikon D7100 camera that is equipped with a 100 mm Tokina lens}. The light source is a camera flash kit (Altura Photo) positioned behind the sample. {We use image analysis technique to extract the particle density profiles and track the position of the sedimentation front over time.}
All experiments are performed at 295 K.

\section{Results and Discussion}

\begin{figure*}[]
\centering
  \includegraphics[width=\textwidth]{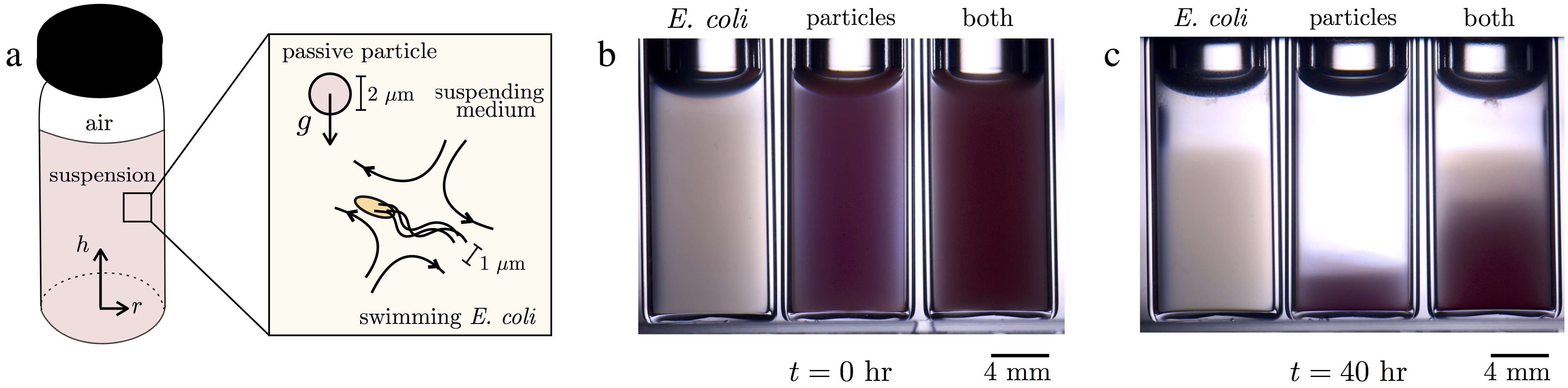}
  \caption{Experimental setup and sample images: (a) A schematic of the setup and bacteria/particle suspensions. Sedimentation experiments are conducted in sealed glass vials that include a volume of atmospheric air. The particles are 2 $\mu$m polystyrene spheres, subject to gravity. The bacteria are 2 $\mu$m rod-shaped \emph{E. coli}, which generate local extensile fluid flows when swimming. Samples are uniformly mixed at the start of the experiments. (b) A sample experiment shows three representative samples: suspensions of (i) only \emph{E. coli} ($c = 1.5 \times 10^9$ cells/mL), (ii) only particles ($1.0 \times 10^8$ particles/mL), and (iii) \emph{E. coli} and particles ($1.0 \times 10^8$ particles/mL, $1.5 \times 10^9$ cells/mL) at $t=0$ hr, the start of the experiment. (c) After 40 hours, the samples have sedimented to various heights. The passive particles sediment much faster than the \emph{E. coli}. When particles and $E.$ $coli$ are combined, the passive particles (pink) extend to higher heights than in the absence of bacteria.}
  \label{Fig_1}
\end{figure*}

Figures \ref{Fig_1}b and c show images of the suspensions taken at $t = 0$ hr (start of the experiment) and $t=40$ hr. The samples in Figure \ref{Fig_1}B and C correspond to, from left to right: (i) a suspension of only $E.$ $coli$ ($c = 1.5 \times 10^9$ cells/mL), (ii) a suspension of only PS particles ($1.0 \times 10^8$ particles/mL), and (iii) a suspension of PS particles and $E.$ $coli$ ($1.0 \times 10^8$ particles/mL, $1.5 \times 10^9$ cells/mL). The samples exhibit a sedimentation front, that is an interface between the aqueous supernatant at the top and the particulate suspension at the bottom. The height of the front depends on the sedimentation velocity of the particles.

A force-balance of gravity and viscous drag yields the bare sedimentation velocity $v_0$ of a single particle sedimenting in fluid of viscosity $\eta$ as $v_0 = \Delta \rho g d^2/{18 \eta}$, where $\Delta \rho$ is the density difference between the particle and suspending solution, $g$ is the gravitational constant ($g = 9.81$ m/s$^2$), and $d$ is the particle diameter. For a two micron polystyrene bead in water, the sedimentation speed is approximately 7 $\mu$m/min. The height of our passive particle suspension is consistent with this bare sedimentation velocity and the dilute concentration of the sample.

The sedimentation process of swimming bacteria significantly differs from passive particles. Indeed, comparison of figure 1b and 1c shows that the bacteria (beige color) settle at a much slower rate than passive particles (pink color) of similar size (2 $\mu$m). This is consistent with the ability of bacteria to self-propel and withstand the external gravitational field in contrast to the passive particles. Interestingly, when we combine swimming bacteria and passive particles in suspension (bottle 3), we find that the sedimentation of the passive particles is hindered: the passive particles (pink) are suspended for longer times at higher heights in the presence of bacteria compared to the passive-particle only case. In this case, the presence of two sedimentation fronts is visible: one front for particles (pink) and one front for bacteria (beige). The front of bacteria is higher than the front of particles and is at the same height as the \emph{E. coli} only case (bottle 1), suggesting that the presence of PS particles does not significantly affect the distribution of bacteria.

\subsection{Sedimentation Profiles}

To estimate particle concentrations along the height $h$ of the bottle, we use image analysis methods to obtain the variations in image intensity $I (x,h)$ throughout the sample; local values of $I (x,h)$ reflect the amount of light that transmits through the sample and decreases with increasing concentration of passive particles and bacteria. We select image intensity profiles as a function of height (SI Fig. 2) from the middle of the bottle, far from the boundaries of the wall to avoid image aberrations. The image intensity profiles are then converted to particle number density through an intensity-density calibration curve, which is determined by measuring the image intensity of suspensions at known concentrations of passive particles and swimming bacteria. The resultant number densities are then multiplied by the volume of the individual particle to obtain the volume fraction as a function of height $h$ (\emph{cf.} Fig. 2 and 4).

In order to characterize the sedimentation process, we plot the normalized concentration profiles $C(h,t)/C_{0}$, where $C_{0}$ is the initial concentration, as a function of distance $h$ (Fig. 2); these profiles are measured at the center of the vial ($x=0$). The normalized concentration profiles are shown at time t=24 hours, t=29 hours, and t=34 hours. Figure 2a shows the case for passive particle only, and the resultant profiles are characterized by a distinct sigmoidal jump which translates in a roughly similar manner as the sedimentation process evolves \cite{martin}. The shape of the concentration profile is consistent with previously measured profiles in passive suspensions of either thermal \cite{Piazza2008} or athermal spherical particles \cite{Davies1988,Lee1992, Salin1994}. The width of of the sedimentation front is related to particle diffusivity, which for small particles in suspension is in part due to thermal motions and in part due to dispersion from long-range hydrodynamic interactions between multiple particles \cite{Ham1988,Chaikin1992,Davies1996,Nicolai1995}. The diffusivity of a sedimenting front of particles can therefore be much greater than Brownian diffusivity.

For suspensions of passive particles, the sedimentation velocity of each particle depends on the position and velocity of the surrounding particles \cite{batchelor, martin}. This leads to a dependence of the sedimenting velocity $U(C)$ on the concentration of particles at that point $C(h,t)$. The concentration $C(h,t)$ is governed by a convection-diffusion equation,
\begin{equation}
	\frac{\partial C}{\partial t} + \frac{\partial (CU(C))}{\partial h} = \frac{\partial}{\partial h}\left(D(C)\frac{\partial C}{\partial h}\right).
	\label{conv_diff}
\end{equation}
If the particle velocity depends linearly on the concentration, (i.e.) $U(C) = \alpha C + \beta$, and the diffusion coefficient $D$ is a constant then the convection-diffusion equation becomes a Burgers' equation that can be solved analytically~\cite{martin}, and the solution is given by:
\begin{equation}
	C(h,t) = \Lambda_{1} + \frac{(\Lambda_{2} -\Lambda_{1})}{1 + \psi(h,t)\exp[(w_{2} - w_{1})(h - V_{S}t)/2D]}. 
	\label{sol_burg}
\end{equation}
where $\Lambda_1=C(0,t)$ and $\Lambda_2=C(L,t)$ are the concentrations of the particles at the boundaries, and $\psi(h,t)=1$ for a steady state profile propagating with a constant drift velocity given by $V_S=(w_1+w_2)/2$. Here $w_{1} = U(\Lambda_{1})$, $w_{2} = U(\Lambda_{2})$ and $U(C)=V_0(1-C)^p\approx V_0(1-pC)$ when $C\sim O(0.1)$. Ignoring the linear correction underestimates the self sharpening characteristic of the sedimentation profile and suppresses the shock formation leading to unusually high estimates of diffusion coefficients \cite{martin}. 

The effect of the linear correction ($pC$) is indeed pronounced when the concentration of particles is large ($C\sim O(0.1)$). However in our experiments the concentration of passive particles is relatively small ($C\sim O(0.001)\ll O(0.1)$), thus the dependence of $U(C)$ on $C$ can be safely ignored and we take $U(C) = V_{S}$ in Eq. (\ref{conv_diff}), which gives similar profiles as Eq. (\ref{sol_burg}).

Figure 2a shows that Eq \ref{conv_diff} with a constant diffusion coefficient $D$ adequately describes the sedimentation profiles of passive particles ($c = 10^8/$mL). The velocity of the sedimentation front $V_{s}$ is approximately $0.12$ $\mu m/s$ which is close to the terminal velocity of a solid sphere of diameter $2$ $\mu m$ obtained from force balance. The fitted diffusion coefficient is $0.75$ $\mu m^2/s$. This value is greater than the thermal diffusivity for a sphere of diameter $d$ of 2 $\mu$m in water at equilibrium, which is approximately 0.2 $\mu m^2$/s as given by the Stokes-Einstein equation $D_0=k_B T/3 \pi \mu d$ \cite{Einstein1905}, where $k_B$ is the Boltzmann constant, $\mu$ is the fluid viscosity, and $T$ is the temperature ($T=295$ K). This is our \textit{control} case. While the effect of passive particles on the sedimentation velocity can be safely ignored (as shown by the fits in Fig. 2a), the effect of active particles can not be. In fact, as we will show below, we find that the effective diffusion coefficient increases and velocity of the resultant sedimentation front decreases due to the presence of live bacteria. We attempt to rationalize these observations in what follows.

\begin{figure*}[]
	\centering
	\begin{subfigure}{.4\textwidth}
		\centering
		\includegraphics[width=0.8\linewidth]{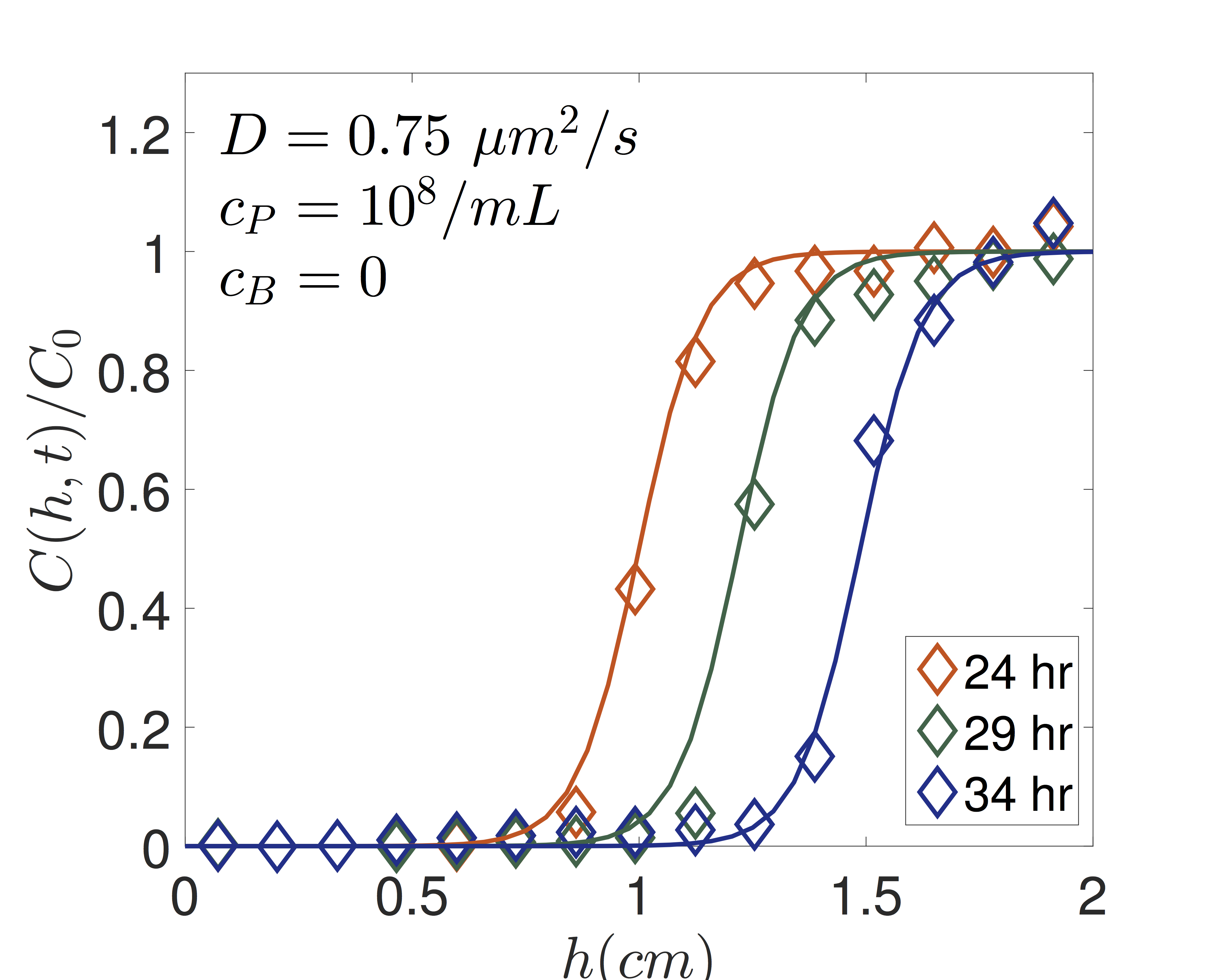}
		\caption{}
		\label{figconc0}
	\end{subfigure}
	\begin{subfigure}{.4\textwidth}
		\centering
		\includegraphics[width=0.8\linewidth]{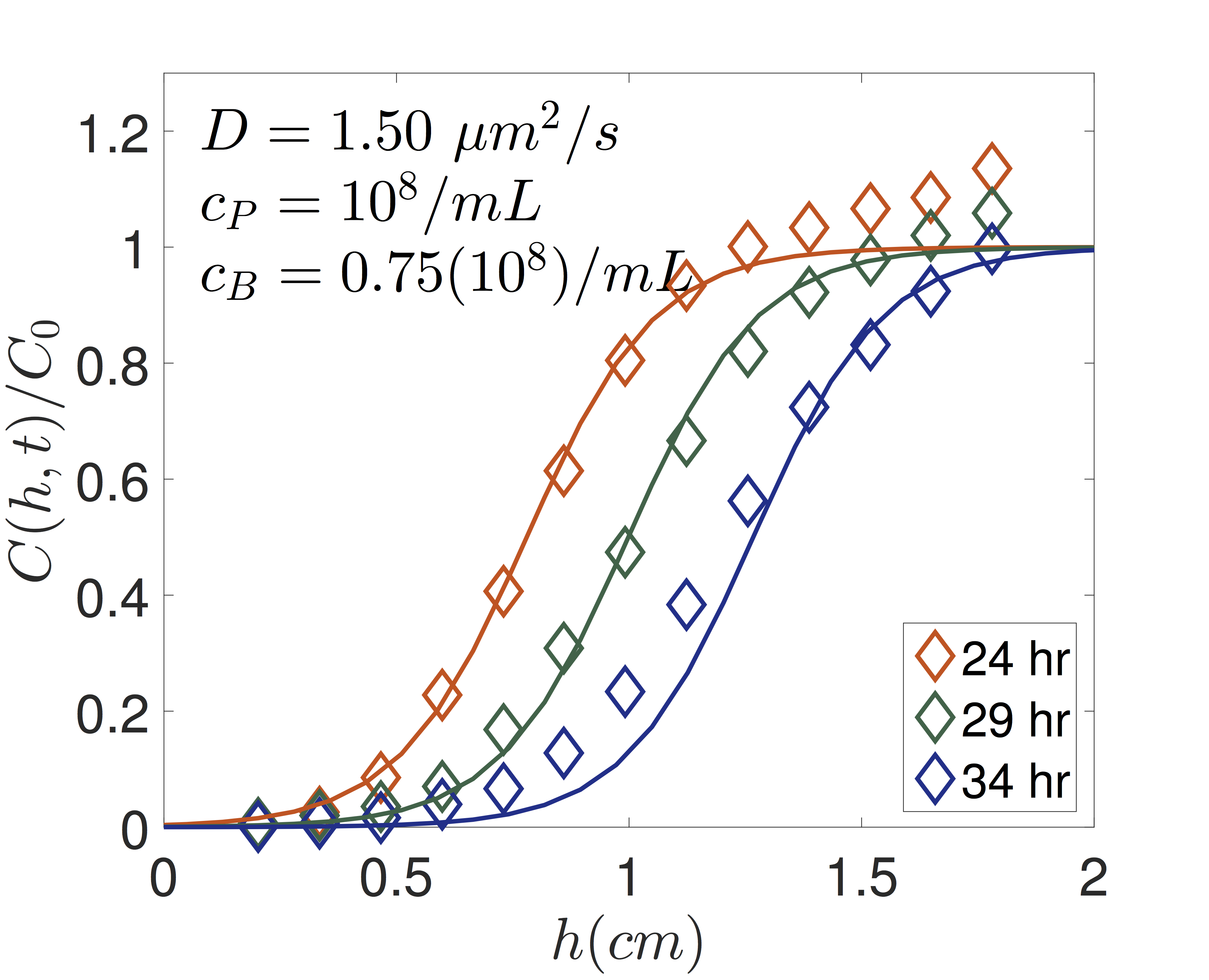}
		\caption{}
		\label{figconc075}
	\end{subfigure}
	\caption{Sedimentation profiles for low concentration of bacteria. Particle suspensions containing (a) no bacteria and (b) low bacteria concentration ($0.75 \times 10^9$ cells/mL).   Suspensions with no or low bacteria concentration can be well described using Burger's equation. Symbols are experimental data and lines are from the solution of Burger's equation. The presence of bacteria increases the diffusion coefficient by $0.75 \mu m^2/s$, however the front propagation velocity $V_S\approx 7$ $\mu$m/min remains almost unchanged. }
	\label{Fig_prof_low}
\end{figure*}

In active suspensions, transport is driven both by diffusion and the motion of live bacteria, which can lead to notable deviations from conventional behavior observed in a diffusion dominated regime~\cite{Patteson2016}. Our experiments indicate that when the concentration of live bacteria is small or comparable to the concentration of passive particles (which is also small) diffusive transport is predominant, thus Eq. (\ref{conv_diff}) with a constant $V_{S}$ and $D$ may still be employed to obtain good fits to the experimental data. For example, Eq. (\ref{sol_burg}) seems to adequately fit the experimental data (Fig. \ref{figconc075}) even when small amounts of bacteria ($0.75\times10^8$ cells/mL) are added to a passive particle suspension ($10^8$/mL). We find that the same value of sedimentation $V_{S}$ is sufficient to capture the density profiles for the case with and without bacteria.

A clear implication of the presence of live bacteria is a marked increase in the effective diffusivity, $D_{\text{eff}}= 1.5~\mu m^2$/s, which is twice the value for the passive particle only case. Studies tracking the microscopic displacements of individual single particles have found that the effective particle diffusivity is the sum of the Stokes-Einstein diffusivity $D_0$ and an active diffusivity $D_\textrm{a}$ that increases linearly with the bacterial concentration \cite{Jepson2013,Kasyap2014,Patteson2016}. Our results on the macroscopic effective diffusion coefficient $D_{\text{eff}}$ during sedimentation show a similar increase due to active contributions \cite{Jepson2013,Kasyap2014,Patteson2016} from swimming bacteria. We further expound upon the impact on the effective diffusion coefficient after presenting the results for higher concentrations of bacteria.

As the concentration of active particles increases while keeping the concentration of passive particles constant at $10^8$/mL, we find significant deviations from the passive particle only case (Fig. 4) and Burgers' equation fails to describe the resultant behavior of the suspension. In an attempt to understand these anomalous observations, we propose a different model that relies on the following two assumptions. First, we assume that the live bacteria in the suspension have a finite life span due to availability of only a finite amount of nutrients and that their death is a first order process. The dead bacteria behave like passive particles and conform to the physics encapsulated in Burgers' equation. Second, the concentration of live bacteria is constant throughout the depth of the bottle $h$, and they die at a constant rate independent of depth and time.

We begin by noting that there are three species of particles in the suspension each of which follows different transport dynamics.
\begin{enumerate}
	\item \textit{Live Bacteria} $(\cl)$: We assume that living \textit{E. coli} are distributed uniformly throughout the bottle and are swimming at speeds ranging from $10-20~\mu$m/s, which is two orders of magnitude larger than the typical magnitudes of terminal velocities of the passive particles ($\sim 0.1 \mu$m/s). In such a scenario, it is reasonable to conjecture that the motion of live bacteria is unlikely to be affected by the motion of the passive particles or the propagation of the sedimentation front (consistent with observations in Fig. 1c). The concentration of live bacteria $C_{1}(h,t)$ depends on time because there is a net death rate of the bacteria which converts them into passive particles. In the simplest case, we model the time varying population of live bacteria using a first order differential equation. 
	\begin{equation}
		\frac{dC_{1}}{dt} = -k C_{1},
	\end{equation}
	where $k$ is a constant. The solution to this equation is 
	\begin{equation}
		C_{1}(h,t) = C_{10}\exp(-kt),
	\end{equation}
	where $C_{10} = C_{1}(h,t=0)$ is the concentration of live bacteria at time $t = 0$. We ignore sedimentation of living bacteria and focus on the  sedimentation of PS particles and dead bacteria.
	\item \textit{Dead Bacteria} $(\cd)$: Death converts bacteria into passive particles that are subject to sedimentation. These new passive particles (dead bacteria) are constantly created at all $h$ and $t$. This results in a source term on the right hand side of our convection-diffusion Eq. \ref{conv_diff} which is given by 
	$-\frac{dC_{1}}{dt}$, so that   
	\begin{equation}
		\frac{\partial C_2}{\partial t} + \frac{\partial(V^b_S C_2)}{\partial h} = \frac{\partial}{\partial h}(D^b\frac{\partial C_2}{\partial h}) + kC_{10}\exp(-kt),\\
		\label{conv_diff_bact}
	\end{equation}		 	
	where $D^b=D^b(C_1)$ and $V^b_S=V^b_S(C_1)$.
	The solution of the partial differential equation above requires two boundary conditions and an initial condition. We apply a no flux boundary condition at the 
	bottom of the bottle $h = 0$ which is:
	\begin{equation}
		D^b\frac{\partial C_2}{\partial h} -V^b_S C_2=0.	
	\end{equation}
	At the top of the bottle we enforce the condition $C_{2}(h=L,t) = 0$. At $t=0$, all the bacteria are alive, hence, the initial condition is $C_{2}(h,t=0) = 0$. 
	\item \textit{Passive Particles} $(\cp)$:
	The third species are the polystyrene spheres whose transport is governed by the usual convection-diffusion equation. 
	\begin{equation}
		\frac{\partial \zeta}{\partial t} + \frac{\partial( V_S^p \zeta)}{\partial h} = \frac{\partial}{\partial h}\left(D^p\frac{\partial \zeta }{\partial h}\right),\\
		\label{conv_diff_pass}
	\end{equation}
	where $V^p_S=V_S^p(C_1)$ and $D^p=D^p(C_1)$.
	We impose a no flux boundary condition at the bottom of the bottle $h=0$:
	\begin{equation}
		D^p\frac{\partial \zeta}{\partial h} -V^p_S \zeta=0.
	\end{equation}
	At the top of the bottle $h=L$, $\zeta(h=L,t)=0$. 
\end{enumerate} 
In the most general case, the diffusion coefficients $D^p$ and $D^b$ as well as the velocities $V_S^p$ and $V_S^b$ in the transport equations given above (Eqs. \ref{conv_diff_bact} and \ref{conv_diff_pass}) could depend both on the concentration of active bacteria and passive particles. 
In fact, in the absence of active particles, the diffusion coefficient $D$ is known to remain constant and the velocity $V_S$ decreases linearly with the concentration of passive particles for dilute suspensions \cite{martin}. Here the concentration of passive particles is small ($<10^{-3}$), and we can ignore even the linear correction to the velocity 
$V_S$ and assume it to be essentially independent of concentration of passive particles $\zeta$. 

\begin{figure*}[]
	\centering
	\begin{subfigure}{.4\textwidth}
		\centering
		\includegraphics[width=0.9\textwidth, height=5 cm]{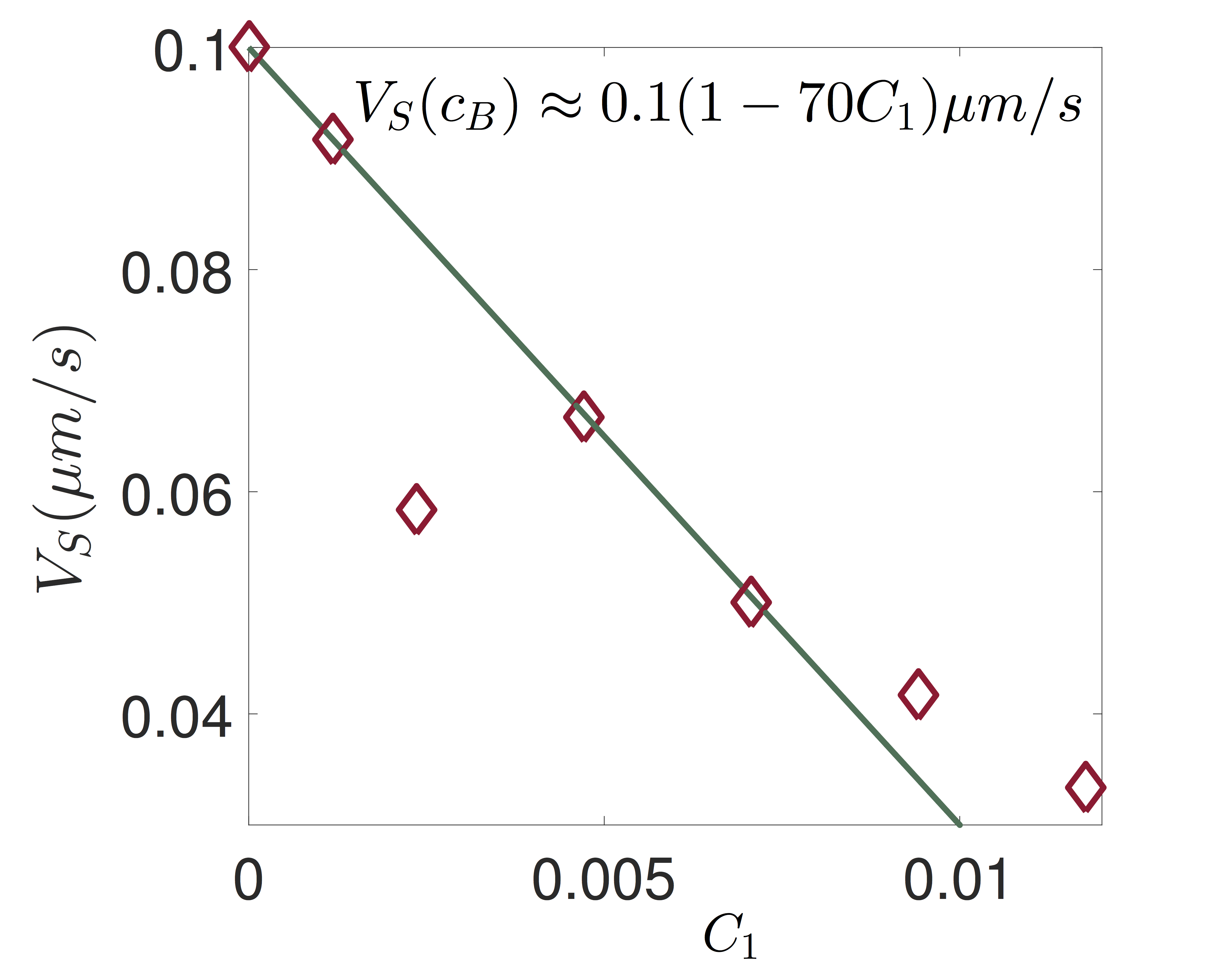}
		\caption{}
		\label{fig_vel}
	\end{subfigure}
	\begin{subfigure}{.4\textwidth}
		\centering
		\includegraphics[width=0.9\textwidth, height=5 cm]{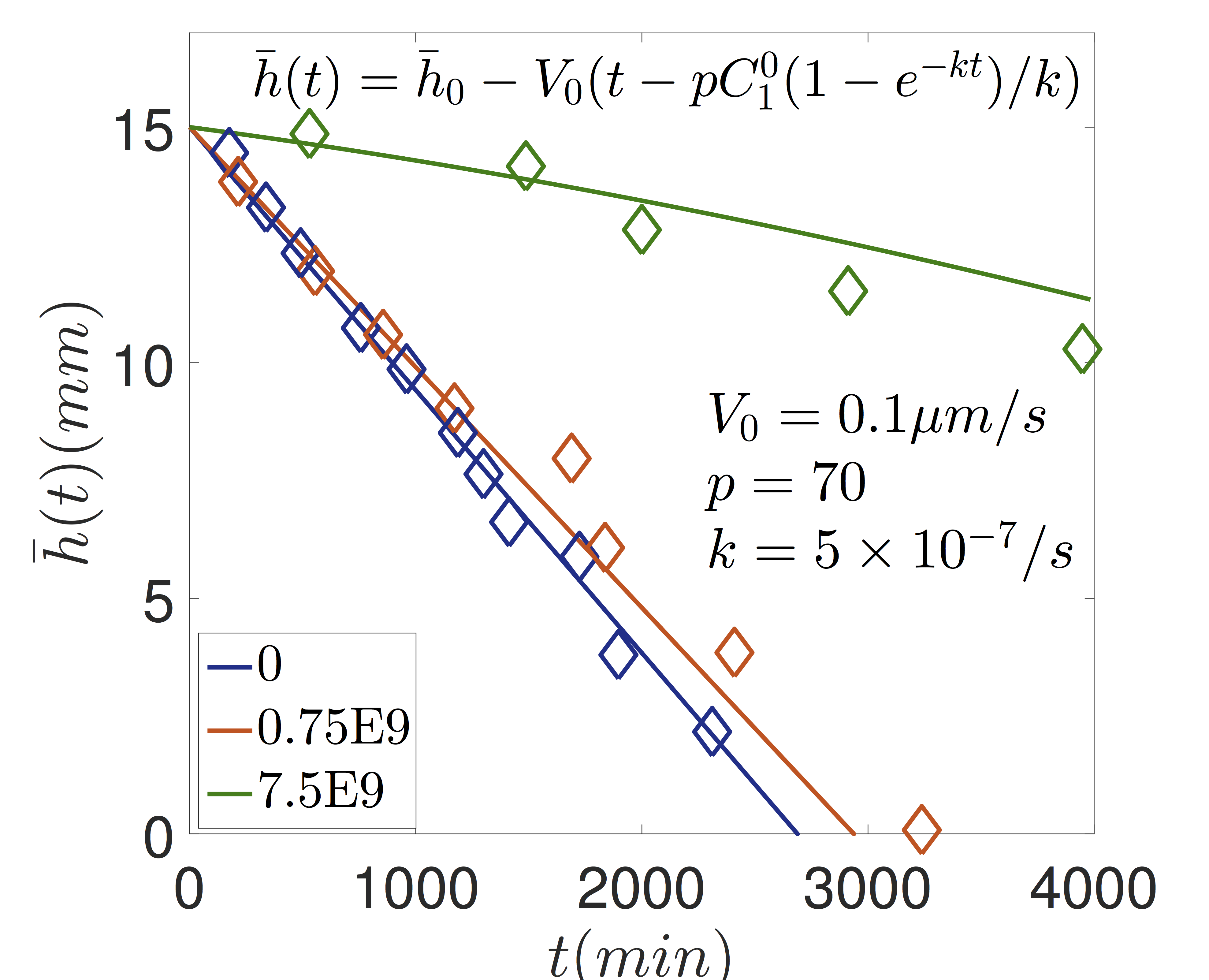}
		\caption{}
		\label{fig_ht}
	\end{subfigure}
	\caption{(a) Sedimentation velocity as a function of concentration of live bacteria, which shows a linear trend for low bacterial concentrations. Experimental data is fitted using the $V(C_1)=V_0(1-pC_1)$ relationship since the data is collected at initial times when $kt\ll1$ and $e^{-kt}\approx 1$ which implies $C_1\approx C_{10}$. Fitting parameters are $p=70$ and $V_0=0.1 \mu m/s$. (b) Position of sedimenting front, $\bar{h}(t)$, as a function of time for passive particles and two bacterial concentrations ($0.75 \times 10^9$ cells/mL and $7.5 \times 10^9$ cells/mL). Note that the expression for the sedimenting velocity in (a) is integrated to obtain the front position such that $\bar{h}(t)=\bar{h}_0-V_0(t-pC_{10}\frac{1-e^{-kt}}{k})$. Using $p$ and $V_0$ values obtained in (a), we are then able to obtain an estimate of $k=5.0 \times 10^{-7}$/s (time scale associated with rate of bacterial death) by fitting to the experimental data in (b).}
	\label{Fig_vel_ht}
\end{figure*}

In the equations above (Eqs. 3-8), the quantities $D^p, D^b, V_S^p, V_S^b$ and $k$ are the unknown parameters that must be estimated by fitting to the experimental data. We start by examining the dependence of front propagation velocity for passive particles $V_S^p$ on the concentration of live bacteria, shown in Fig \ref{fig_vel}. The data shows that the sedimentation front velocity $V_S$ decreases nearly linearly as the the concentration of live bacteria $C_1$ increases. We assume a relationship of the form $V_S^p(C_1)=V_0^p(1-C_1)^p\approx V_0^p(1-pC_1)$ when $C_1\ll1$.  By fitting to the experimental data in Fig. \ref{fig_vel}, we find $V_0^p =V_S^p(C_1=0)\approx 0.1~\mu m/s$ and $p=70$. As expected, the quantity $V_0^p$ is approximately equal to the the front propagation velocity in the absence of any bacteria. A dramatic deviation is, however, observed in the exponent $p(\sim70)$, which usually ranges from 4.5 to 6 for passive particles \cite{martin,batchelor}. This unusually high value of the exponent $p$ highlights the role of activity in sedimenting suspensions.

Next, we assume the same form for the sedimentation of dead bacteria, \textit{i.e.} $V_S^b=V_0^b(1-pC_1)$. The shape of \textit{E. Coli} is similar to a cylinder with length $1~\mu m$ and diameter $2~\mu m$, thus it experiences an anisotropic drag. For the sake of simplicity, we assume \textit{E. Coli} to be spheres with effective diameter of $d^b=1.44~\mu m$. The difference in density for \textit{E. coli} and surrounding solution $\Delta \rho^b$ is assumed to be similar to the difference in density for polystyrene and the solution $\Delta \rho^p$, and the terminal velocity of a bacterium is then proportional to the square of the effective diameter. Thus, we thus obtain $\frac{V_0^b}{V_0^p}=(\frac{d^p}{d^b})^2\approx \frac{1}{2}$ which implies $V_0^b=0.06\mu m/s$. Since $V^{b}_S(C_1(t))=V^{b}_S(t)$ is a function of time $t$ only, $\frac{\partial (C_2 V_S^{b})}{\partial h}=V_S^b\frac{\partial C_2}{\partial h}$ in Eq.(\ref{conv_diff_bact}). Similarly, $V^p_S$ is also devoid of any spatial gradients and $\frac{\partial (\zeta V_S^{p})}{\partial h}=V_S^p\frac{\partial \zeta}{\partial h}$ in Eq  (\ref{conv_diff_pass}).

Our next step is to obtain the value of $k$, rate of bacterial death, which plays a key role in the population dynamics of bacteria. In order to do so, we observe the sedimentation front for passive particles over a substantial period of time ($15-50$ hours) for different initial concentrations of live bacteria. From our previous discussion, $V_S^p=V_0(1-pC_1)=V_0(1-pC_{10}e^{-kt})$, which enables us to obtain the height of the front $\bar{h}(t)$ as function of time. 
\begin{equation}
	\bar{h}(t)=\int_{0}^{t}V_S^p~dt=\bar{h}_0-V_0(t-pC_{10}\frac{1-e^{-kt}}{k}).
\end{equation}          
We substitute the values of $p$, $V_0$ and $C_{10}$ and are left with only one fitting parameter $k$. Fitting to the experimental data (Fig. 3b), we obtain $k=5.0\times10^{-7}/s$. We note that increasing $k$ by 10 times does not have much effect on the profiles of $\bar{h}(t)$. Finally, for simplicity, we assume $D^p=D^b=D(C_1)$. Here, we point out that changing the diffusion coefficient by some amount ($\sim 10\%$) does not have any noticeable effect on the profiles. Finally, we note that a typical experiment runs for almost $40$ hours, thus $kt<5\times10^{-7}\times40\times3600\approx0.072$ which gives $0.92<e^{-kt}<1$. Thus, for a given bottle with a fixed initial concentration of live bacteria $C_{10}$, $D(C_1)=D(C_{10}e^{-kt})\approx D(C_{10})$, as such the diffusion coefficient depends only on the initial concentration of live bacteria. 

\begin{figure*}[]
	\centering
		\begin{subfigure}{.4\textwidth}
			\centering
			\includegraphics[width=0.9\textwidth, height=5 cm]{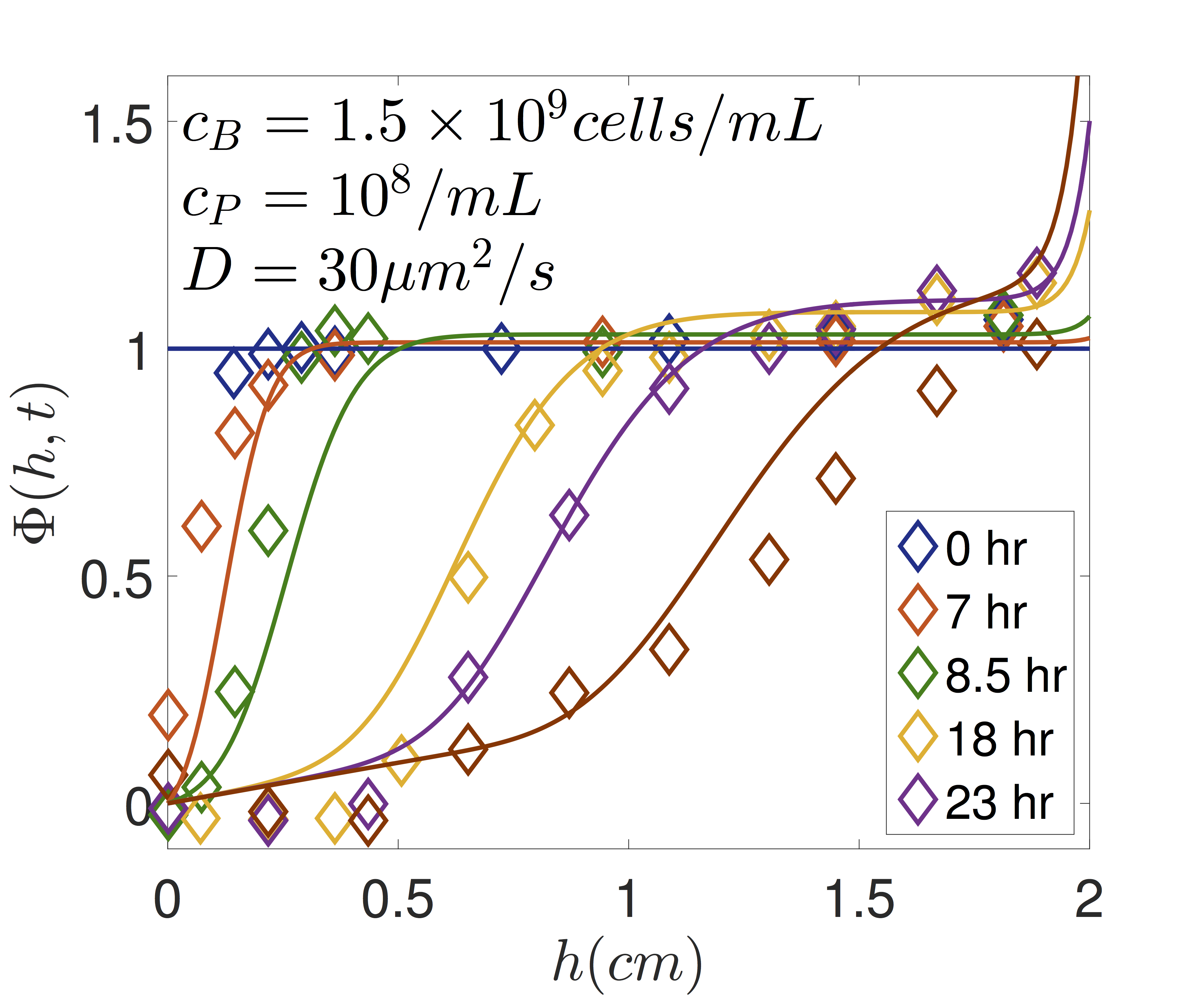}
			\caption{}
		\end{subfigure}
		\begin{subfigure}{.4\textwidth}
			\centering
			\includegraphics[width=.9\textwidth, height=5 cm]{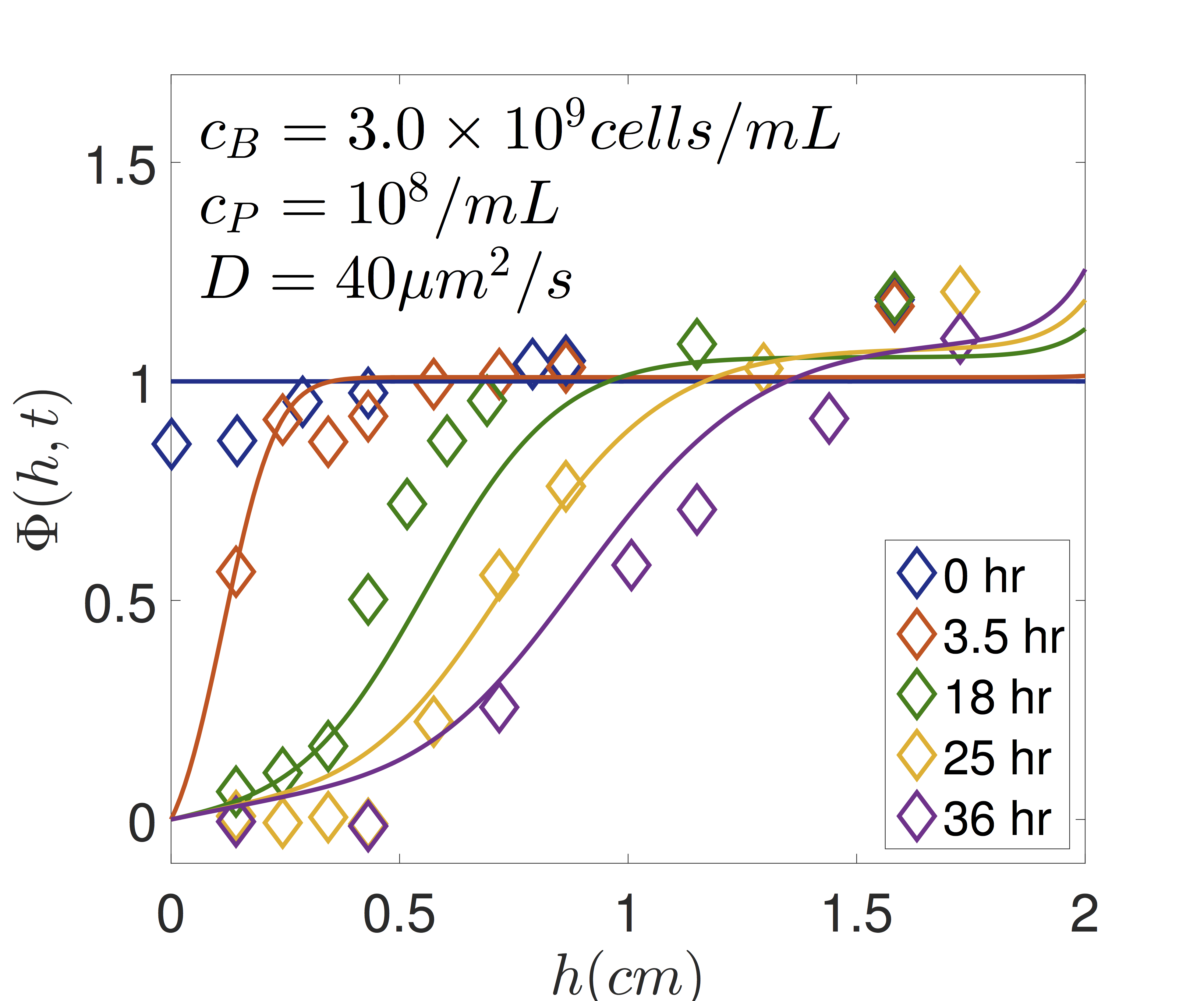}
			\caption{}
		\end{subfigure}
		\begin{subfigure}{.4\textwidth}
			\centering
			\includegraphics[width=.9\textwidth, height=5 cm]{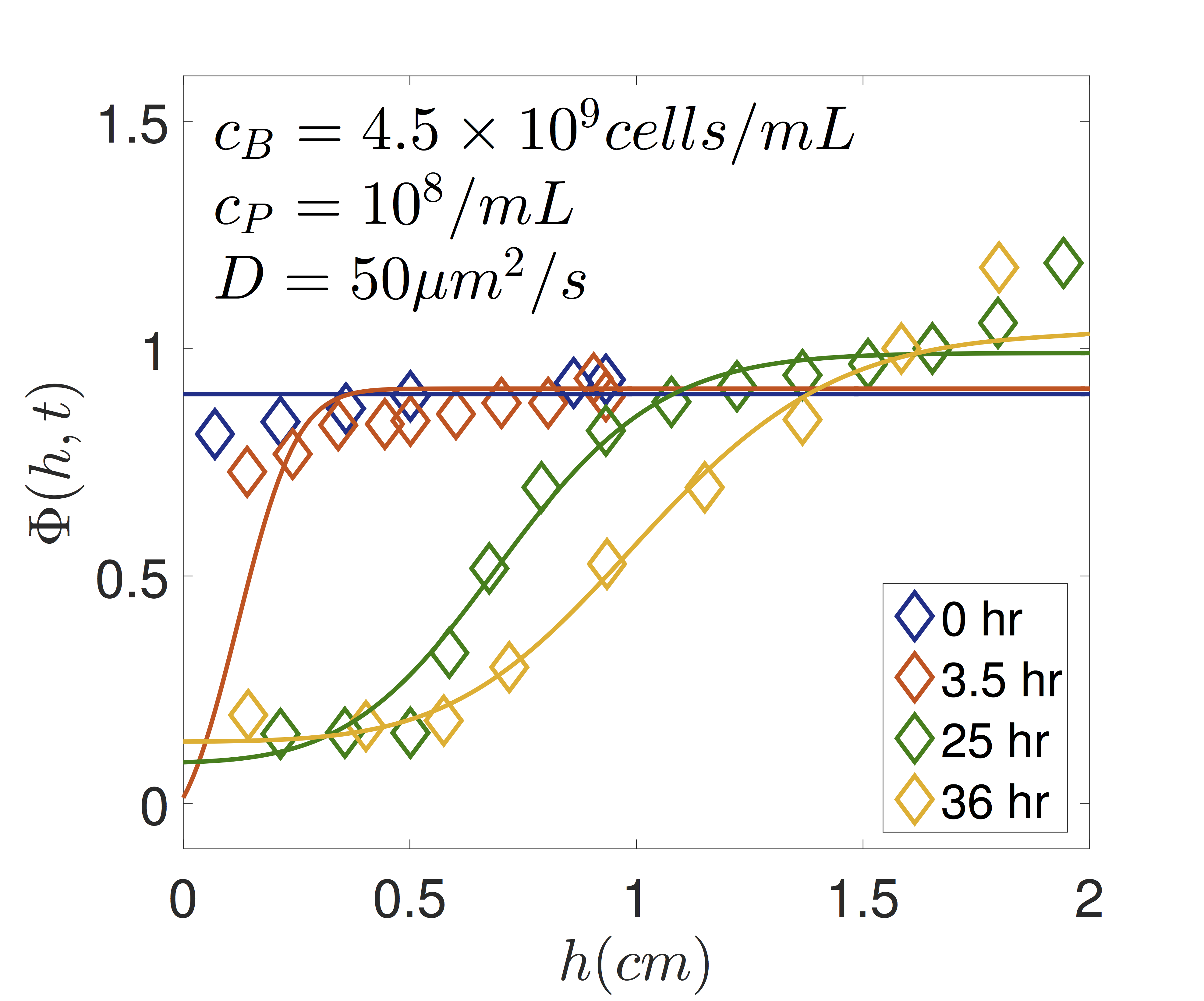}
			\caption{}
		\end{subfigure}
		\begin{subfigure}{.4\textwidth}
			\centering
			\includegraphics[width=.9\textwidth, height=5 cm]{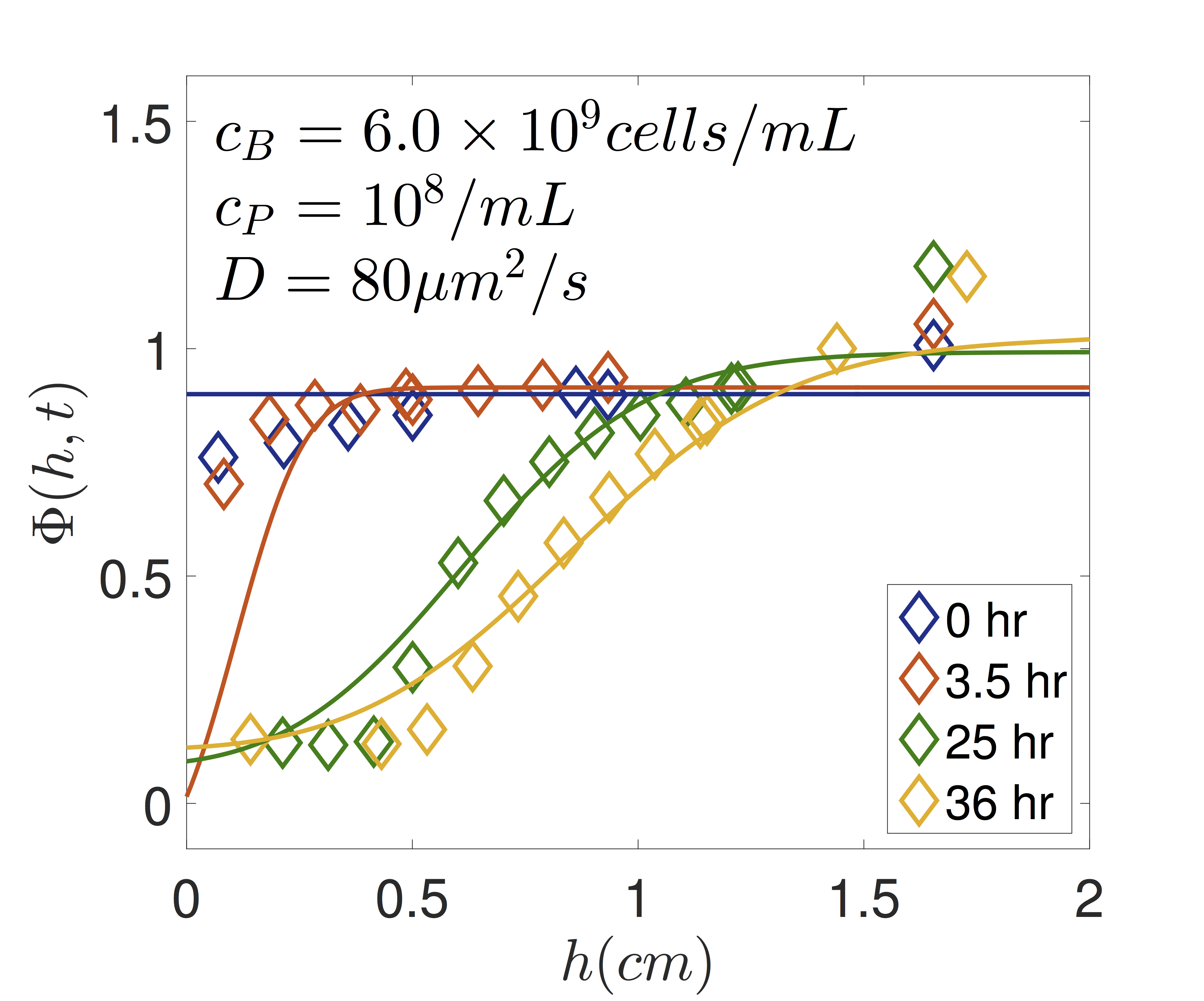}
			\caption{}
		\end{subfigure}
	\caption{Sedimentation profiles for the cases of higher bacterial concentrations. (a) $1.5 \times 10^9$ cells/mL, (b) $3.0 \times 10^9$ cells/mL, (c) $4.5 \times 10^9$ cells/mL, and  (d) $6.0 \times 10^9$ cells/mL  Note that the diffusion coefficients increases as the concentration of bacteria increases and are larger compared to the ones obtained in Fig \ref{Fig_prof_low}.  Here $\Phi(h,t)=r_B\frac{C_2(h,t)}{C_0}+r_P\frac{\zeta(h,t)}{\zeta_0}$, $r_P=0.9,r_B=0.1$. We use the linear variation of sedimentation velocity with the concentration of live bacteria to obtain these profiles by integrating Eqns. (\ref{conv_diff_bact}) and (\ref{conv_diff_pass}).}
	\label{Fig_prof_high}
\end{figure*}

We now integrate the partial differential equations (Eqs. \ref{conv_diff_bact} and \ref{conv_diff_pass}) along with the associated boundary conditions to obtain $C_2(h,t)$ and $\zeta(h,t)$. Since, the live bacteria are almost transparent, the aggregate intensity $I(h,t)$ observed in the bottle is proportional to concentrations of passive particles weighted by their relative opacities \textit{i.e.} $I(h,t)\propto r_bC_2(h,t)+r_p\zeta(h,t)$, where $r_p$ and $r_b$ denote the opacities of polystyrene particles and dead bacteria, respectively. The resultant sedimentation
profiles are shown in Fig. \ref{Fig_prof_high} together with experimental data. In each panel the concentration of passive particles is held fixed at $\zeta = 10^{8}$ while the concentration of live bacteria
increases from panel 4a to panel 4d. The lines obtained from integration of Eq. (\ref{conv_diff_bact}) and (\ref{conv_diff_pass}) match the profiles reasonably well including capturing the accumulation of
particles at the bottom of the bottles due to the no flux boundary conditions. 

\begin{figure*}[]
	\centering
			\begin{subfigure}{.4\textwidth}
				\centering
				\includegraphics[width=.9\textwidth, height=5 cm]{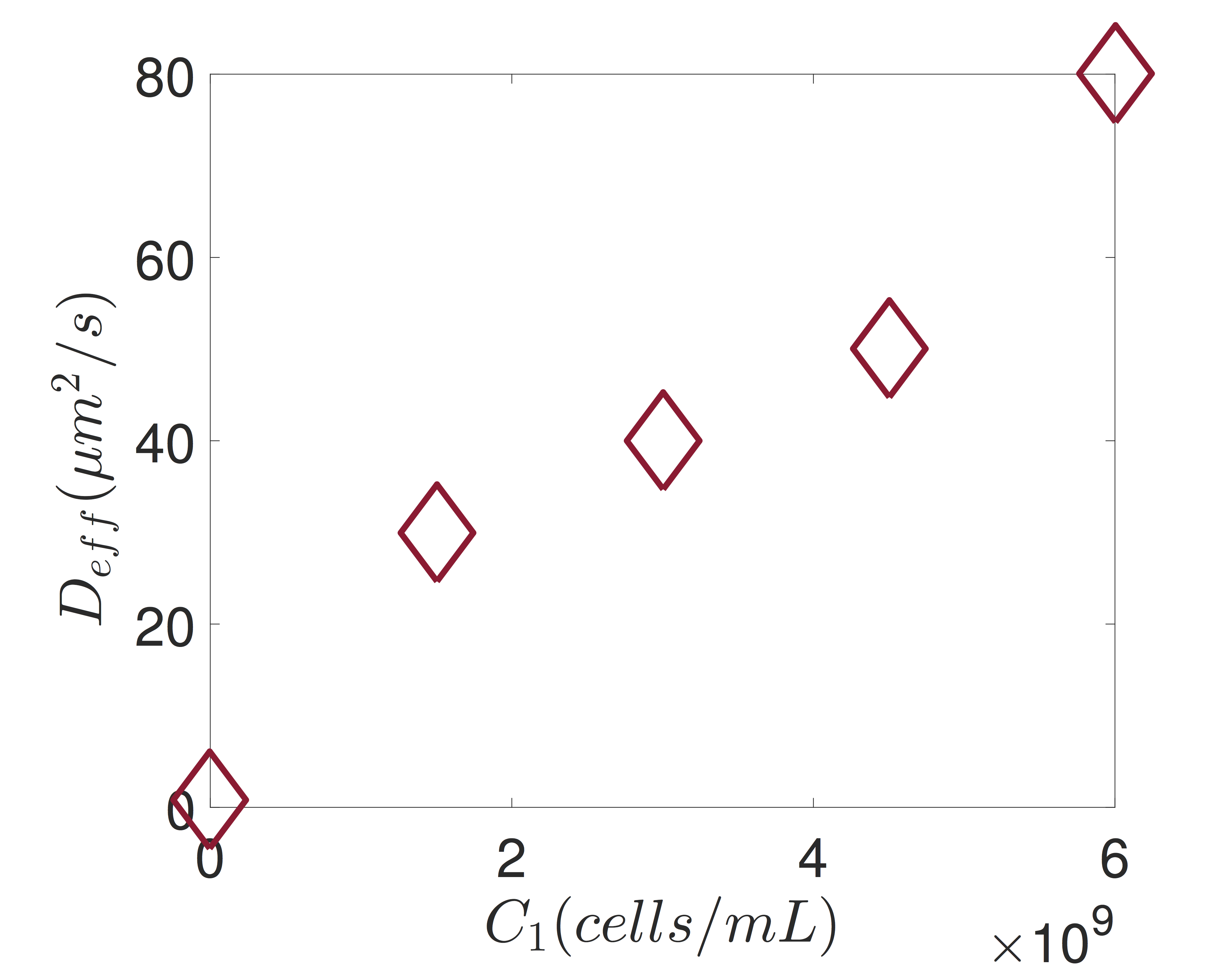}
				\caption{}
			\end{subfigure}		
			\begin{subfigure}{.4\textwidth}
			\centering
			\includegraphics[width=.9\textwidth, height=5 cm]{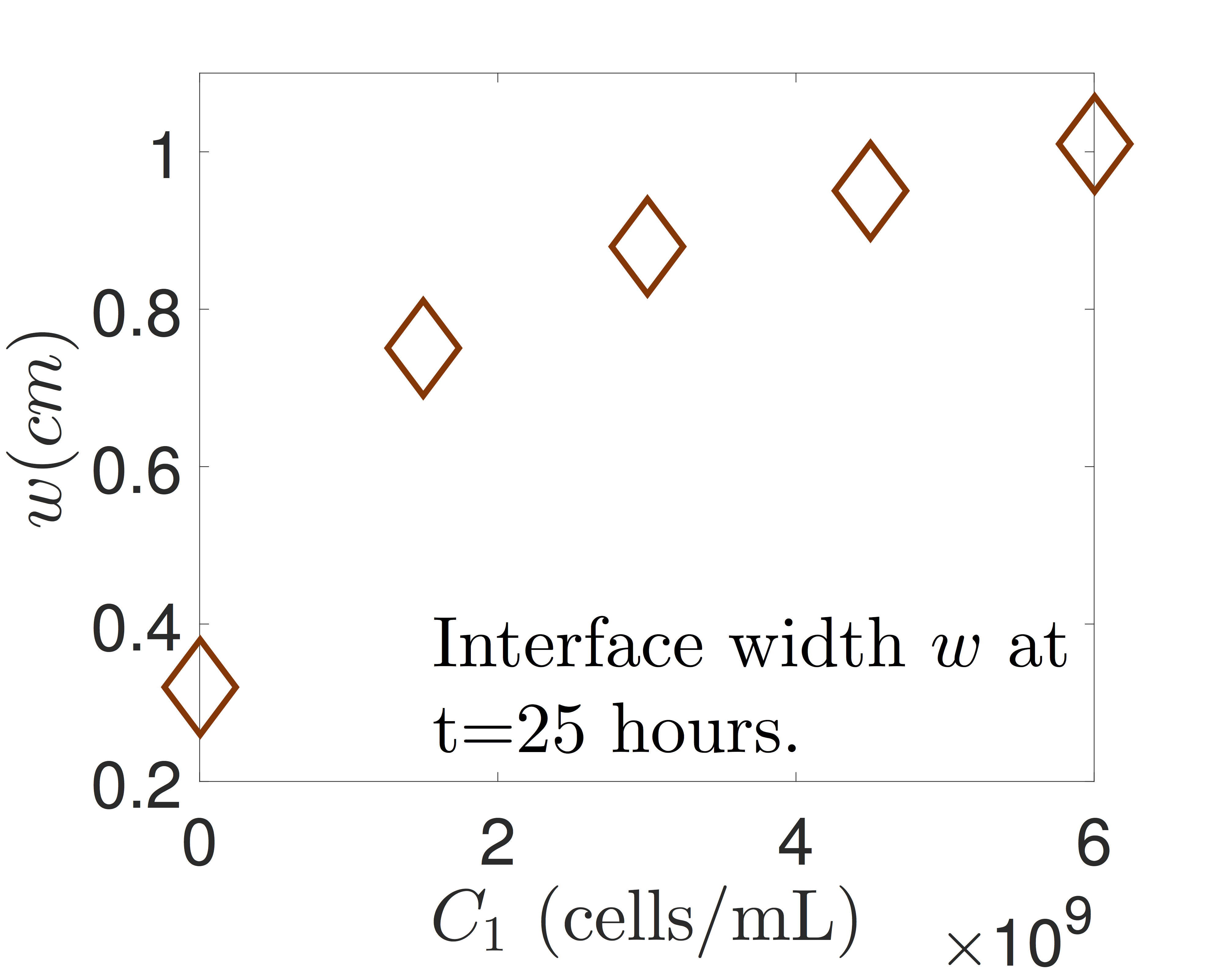}
			\caption{}
		\end{subfigure}
	\caption{(a) Effective diffusion coefficient $D_\textrm{eff}$ as a function of bacterial concentration. Here, we assumed $D^b=D^p=D_\textrm{eff}$. (b) The width of the sedimentation front increases as the bacteria concentration increases.}
	\label{Fig_diff}
\end{figure*}

We find that the presence of live bacteria in sedimenting suspensions impacts the behavior of the suspension in two profound ways: we find that (i) the velocity of the sedimentation front decreases with increasing concentration of live bacteria and (ii) the diffusion coefficients, in case of suspensions containing large concentration of live bacteria (Fig. \ref{Fig_prof_high}), are much larger ($\sim10$ times) than those observed in cases where bacteria are either absent or present in small concentrations (Fig \ref{Fig_prof_low}). The variation of the fitted diffusion coefficient with live bacteria concentration is plotted in Fig. \ref{Fig_diff}a. The diffusion coefficient increases with the concentration of live bacteria, consistent with the corresponding increase in the width of the sedimentation front (Fig. \ref{Fig_diff}b). Previous investigations \cite{Mino2011,Jepson2013,Kasyap2014,Patteson2016} have measured particle diffusivity as a function of bacterial concentration by tracking the microscopic displacements of individual particles over time; these reports find that at long times the particles behave diffusively with a diffusion coefficient that increases linearly as the concentration of living bacteria increases \cite{Mino2011,Jepson2013,Kasyap2014,Patteson2016}. In contrast to this self-diffusion process, in our experiments, we observe the gradient diffusion of passive particles, which is due to motion of fluctuating particles in response to a concentration gradient \cite{Davies1988}. Here, we find that the particle diffusivity increases monotonically with increasing bacteria concentration but at a rate that is significantly larger (by a factor of approximately 40) than those measured in previous studies for particle self-diffusion in suspensions of \emph{E. coli} \cite{Patteson2016}. Taken together, we reason that the significant increase in particle diffusion and drastic drop in sedimentation velocity is in part due to enhanced particle transport due to steric and hydrodynamic interactions with swimming bacteria at the micro-scale \cite{Kasyap2014,Patteson2016} and may in part be due to bacteria-mediated changes in fluctuations and long-range hydrodynamic interactions among particles.

\subsection{Sedimentation Fluctuations}

\begin{figure*}[]
\centering
  \includegraphics[width=.8\textwidth]{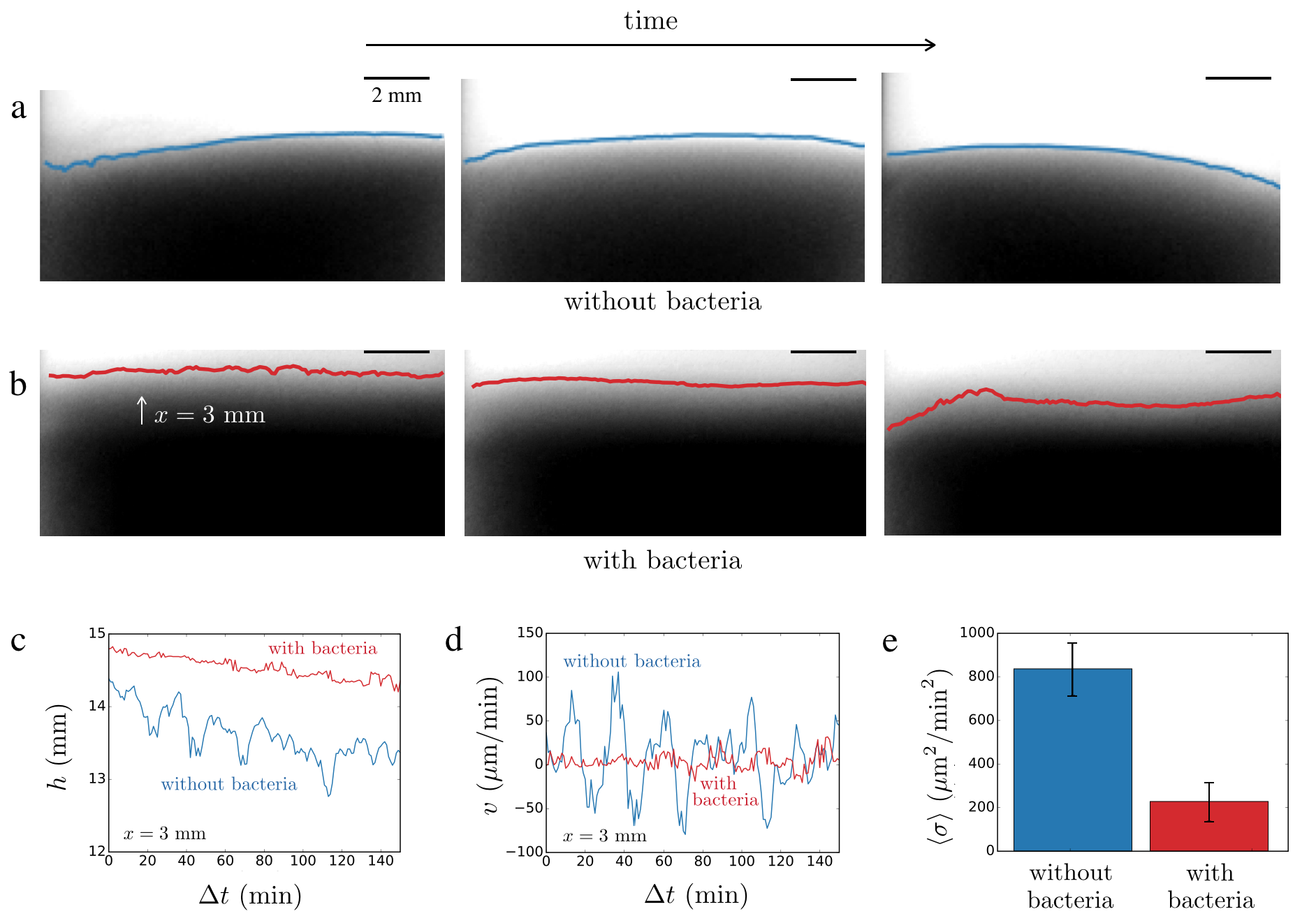}
  \caption{The presence of bacteria suppresses fluctuations in the sedimentation front. (a \& b) Sample images of the sedimentation front in the case without bacteria and the case with bacteria ($c=1.5 \times 10^9$ cells/mL) shown at 20 minutes intervals. The images are overlaid by traces of the tracked position of the front $h(x)$ (blue and red lines). (c \& d) Representative plots of the front position and front velocity versus time, $\Delta t$. Here $\Delta t=t-t_0$, where $t_0 =$300 minutes, when the sedimentation front or the particles are distinct from the fluid-air interface. Data in (c) \& (d) was obtained a distance $x = 3$ mm from the edge of the bottle. (e) The root mean square velocity $\langle \sigma \rangle$ significantly decreases when bacteria are added to the suspension. }
  \label{Fig_4}
\end{figure*}

Having shown that that presence of bacteria can significantly affect the mean sedimentation velocity of a suspension of passive particles, we next examine the role of swimming \emph{E. coli} in fluctuations of the sedimenting front \cite{Hinch2010}. We track the spatially-varying position of the sedimentation front as a function of time (Fig. 6). The front position $h(x)$ is obtained at each time by identifying the inflection point in the density versus height curves (Fig. 2 and 3) for varying positions $x$ across the bottle, as shown at 20 min. intervals in Fig. \ref{Fig_4} for the case without bacteria and the case with bacteria ($c=1.5 \times 10^9$ cells/mL). Over this time interval, the mean position of the fronts has not changed significantly, but fluctuations about the mean position are evident. For the case without bacteria, the front exhibits fluctuations that span the width of the bottle (approximately 8 mm). With bacteria, the front appears to fluctuate over smaller length scales (1-2 mm) and has a smaller amplitude than the case with bacteria.

To quantify the temporal fluctuations of the sedimentation front, we track the height and velocity of the sedimentation front at varying positions across the width of the bottle over time (Fig. \ref{Fig_4}c,d). Here, we define local velocities of the sedimentation front as $(h(x,t_0)-h(x,t_0+\Delta t))/\Delta t$, where $h(x,t)$ is the position of the sedimentation front at time $t$ and $\Delta t = 8$ minutes. Figure \ref{Fig_4}c and d show representative data of the sedimentation front height $h$ and velocity $v$ at a distance $x=3$ mm (Fig. \ref{Fig_4}) from the wall of the bottle. We find that the passive suspension of dilute 2 $\mu$m particles undergoes velocity fluctuations up to a maximum of 100 $\mu$m/min. When bacteria are added to the solution ($c = 1.5 \times 10^9$ cells/mL), the fluctuations significantly decrease, with a maximum less than 20 $\mu$m/min. The root mean square velocity ($\langle \sigma \rangle$) of the sedimentation front is four times smaller (Fig. ~\ref{Fig_4}e) in the presence of bacteria.

Taken together, Fig. \ref{Fig_4} demonstrates that fluctuations in the sedimentation front are significantly smaller in the presence of bacteria compared to the case without bacteria. The decrease in particle fluctuations in the presence of bacteria may in part stem from the enhancement in particle diffusivity. In the sedimentation of particles in passive fluids, local gradients in concentration, such as at the spreading sedimentation front, drive fluctuations \cite{Hinch2010}. Here, we observe that fluctuations are higher in the absence of bacteria, where particle concentration gradients are relatively sharp, and lower in the presence of bacteria, where concentration gradients are relatively smaller (Fig. 1, 2, and 3). We postulate therefore that the decrease in fluctuations is due to bacteria-mediated spreading of particles, which smooths out large concentration gradients.

\section{Conclusions}
The sedimentation of passive particles in the presence of live bacteria is investigated both in experiments and using a simple model. We find that the presence of swimming bacteria can significantly impact the sedimentation process of passive particles. At low concentrations of live bacteria ($c_b < 1 \times 10^9$ cells/ml), we find that the presence of bacteria increases the diffusion coefficient of the passive particles, while the mean sedimentation velocity remains constant. As the concentration of bacteria $c_b$ increases, we observe strong deviations from this behavior: the diffusion coefficient of the passive particles increases with $c_b$ (Fig \ref{Fig_diff}a) and the sedimentation velocity decreases rapidly compared to passive particle suspensions, even for concentrations of particles and bacteria considered dilute (volume fraction $<$ 1\%) (Fig \ref{Fig_vel_ht}a). Our model shows that the assumption that the death of bacteria over the course of the experiment, which creates a source of passive particles, is necessary to capture the experimental data. Thus, the description of sedimentation in terms of the classic Burgers' equation is not sufficient for describing the processes occurring in the presence of bacteria. However, Burgers' equation can be modified to yield a reasonable model for sedimentation. The key modifications to Burgers' equation are that (a) the diffusion coefficient entering the right hand side of Burgers' equation is a function of live bacteria concentration and varies with time, (b) the particle velocity on the left hand side of the  equation is a function of live bacteria concentration that also varies with time, and (c) a time dependent source of passive particles also appears in the governing equation due to death of bacteria.

Surprisingly, we find that the presence of bacteria decreases the root-mean-square fluctuations of the sedimentation front when compared to the case of purely passive particles. The suppression in fluctuations may be due to the increase in the sample effective diffusivity due to the swimming motion of \emph{E. coli}, which may be smooth out large concentration gradient that would otherwise exist in the absence of bacteria.

Our study has implications for describing the sedimentation process in which active particles are present. We have shown that, in describing such active systems, population dynamics of bacteria cannot be ignored. Here, we have treated the population dynamics in a simple manner and shown that it was sufficient to account for the observations in experiment. Our treatment was sufficient likely because the bacteria were isolated in the bottles once the experiments began with no access to nutrients except oxygen. However, more sophisticated treatments might be necessary in more complex situations.  More broadly, our study could have implications on sedimentation processes in geological and man-made water reservoirs in which live micro-organisms are almost always present. A natural next step would be to explore the role of the particle size in sedimentation, since larger particles can diffuse faster than smaller particles in suspensions of swimming bacteria \cite{Patteson2016}; this effect could lead to anomalous sedimentation velocities and diffusion coefficients, which may control particle sorting during sedimentation. It would also be interesting to explore how extracellular polymers, which bacteria excrete and which can modify microbial swimming behavior \cite{PattesonSciRep,Qin2015}, impact particle sedimentation and spatial organization in active environments.

\section*{Acknowledgements}
We kindly thank Arvind Gopinath, Howard Stone, Aparna Baskaran, and Ravi Radhakrishnan for fruitful discussions. A. E. Patteson and P. E. Arratia acknowledge support by the National Science Foundation grant CBET-1437482. J. Singh and P. K. Purohit acknowledge support through National Science Foundation grant DMR-1505662 and the University Research Foundation at the University of Pennsylvania. A. E. Patteson was supported by NSF Graduate Research Fellowship.


\newpage{}



\end{document}